\documentclass[aps,prd,preprint,groupedaddress]{revtex4-1}
\usepackage[mathscr]{euscript}
\usepackage{graphicx}
\usepackage{float,epsfig}
\usepackage{bm}
\usepackage{graphicx}
\usepackage{subfigure}
\usepackage[colorlinks=true,linkcolor=blue]{hyperref}
\newcommand{\bea}{\begin{eqnarray}}
\newcommand{\eea}{\end{eqnarray}}
\newcommand{\beq}{\begin{equation}}
\newcommand{\eeq}{\end{equation}}

\def\/{\over}

\begin{document}


\title{Resonance interaction between two entangled  gravitational polarizable objects}

\author{Yongshun Hu$^{1}$, Jiawei Hu$^{1}$\footnote{jwhu@hunnu.edu.cn}, Hongwei Yu$^{1}$\footnote{hwyu@hunnu.edu.cn} and Puxun Wu$^{1}$\footnote{pxwu@hunnu.edu.cn}  }

 \affiliation{$^{1}$Department of Physics and Synergetic Innovation Center for Quantum Effects and Applications, Hunan Normal University, Changsha, Hunan 410081, China 
}


\begin{abstract}

We investigate the resonance quadrupole-quadrupole interaction between two entangled gravitationally polarizable objects induced by a bath of fluctuating quantum gravitational fields in vacuum in the framework of linearized quantum gravity.  Our result shows that, the interaction energy behaves as $r^{-5}$ in the near regime, and oscillates with a decreasing amplitude proportional to $r^{-1}$ in the far regime, where $r$ is the distance between the two objects. Compared to the case when the two objects are in their ground states, the quantum gravitational interaction is significantly enhanced when the objects are in an entangled state. Remarkably, in the far regime, the resonance quantum gravitational interaction can give the dominating quantum correction to the Newtonian potential,  since the extremum is much greater than the monopole-monopole quantum gravitational interaction.

\end{abstract}


 \maketitle

\section{Introduction}
\label{sec_in}
\setcounter{equation}{0}
Gravitational waves, which are ripples of spacetime predicted by Einstein based on his  general theory of relativity~\cite{ES}, have been directly detected recently by the Laser Interferometer Gravitational-Wave Observatory (LIGO)~\cite{BPA}. It is well-known that gravitational waves may cause  length difference between the arms of a laser interferometer, which is a classical effect of gravitational waves detected by LIGO. Naturally, one may wonder what are the quantum effects of quantized gravitational waves, and whether such effects are detectable. Unfortunately, a full theory of quantum gravity is  elusive at present. Even though, one can still study quantum gravitational effects at low energies in the framework of linearized quantum gravity. One such example is the quantum light-cone fluctuations~\cite{HL, HLF, YuWu, HNF}. Also, it has been found that, by summing one-loop Feynman diagrams with off-shell gravitons where general relativity is treated as an effective field theory at low energies, the interaction between two mass monopoles shows a quantum correction to the Newtonian force, which behaves as $r^{-3}$~\cite{JF}.

Quantum mechanically, there inevitably exist quantum vacuum fluctuations of gravitational fields when gravity is quantized, which induce instantaneous quadrupole moments in gravitationally polarizable objects. In analogy to the electromagnetic Casimir-Polder (CP) effect~\cite{CP}, there should also be gravitational CP effect between a gravitationally polarizable object and a gravitational boundary~\cite{JH}, and between two gravitationally polarizable objects~\cite{LMJ,BJ,PJ,PJH,HZW}. The CP-like quantum gravitational force between a pair of polarizable objects in their ground states has been shown to  be proportional to $r^{-10}$ and $r^{-11}$ in the near and far regimes respectively~\cite{LMJ,BJ,PJ,PJH}.  Thus, a question naturally arises as to whether such quantum gravitational effects can be more significant in certain circumstances. Fortunately, there are similar examples in quantum electrodynamics. The quantum interaction between two electrically neutral atoms behaves as $r^{-7}$ in the far-zone limit $r\gg\lambdabar$ (where $\lambdabar$ is the reduced characteristic transition wavelength) when both of them are in their ground states~\cite{CP}, while it may be much greater  when one or both  atoms are in an excited state and an exchange of real photons is involved, which is referred to as resonance interaction~\cite{PT,Salam}.  For example, the resonance interaction  may vary as $r^{-2}$ when one of the two electrically neutral atoms is prepared in the ground state and the other in the excited state~\cite{ET,AT,LRF,PRB,MRA,PRLS,PS}, and $r^{-1}$ when the two atoms are in a symmetric/antisymmetric entangled state~\cite{PT}, which are much greater than the $r^{-7}$ CP potential.  Therefore, one may expect that the quantum gravitational potential may also be much greater when two objects are prepared in an entangled state.

In this paper,  we  investigate the quantum gravitational interaction between two entangled quantum objects coupled with a bath of fluctuating quantum gravitational field in vacuum in the framework of linearized quantum gravity with the method proposed by Dalibard, Dupont-Roc, and Cohen-Tannoudji (DDC)~\cite{JJ,JJC}, which has recently been  exploited to study the  effects of spacetime curvature and topology on the resonance interaction energy between two entangled electrically neutral atoms  \cite{WenYu,WRL1,WRL2,WLR,LM,rie17}.
Firstly, we derive the formulae for the effective Hamiltonian of the two objects. Then, we calculate the interaction energy  between the two objects based on the theory of linearized  quantum gravity. In this paper, the Latin indices run from $1$ to $3$, the Greek indices run from $0$ to $3$, and the Einstein summation convention is assumed. Natural units with $c=\hbar=16\pi G=1$ are used, where $c$ is the  speed of light, $\hbar$ the reduced Planck constant, and  $G$  the Newtonian gravitational constant.

\section{Basic equations}
\label{sec_ba}
The system we consider consists of two entangled two-level objects in interaction with a bath of fluctuating  gravitational fields in vacuum. We assume that the two objects (labelled as A and B) can be treated as harmonic oscillators with two internal energy levels, $\pm\frac{1}{2} \omega_{0}$, associated with the eigenstates $ |e\rangle$ and $|g\rangle $, respectively. The total Hamiltonian takes the form
\beq
H=H_{F}+H_{S}+H_{I},
\eeq
where $H_{F}$ is the Hamiltonian of the gravitational field, and $H_{S}$ is the Hamiltonian of the two-level systems (A and B), which can be given as
\beq
H_{S}=\omega_{0}\sigma^{A}_{3}+\omega_{0}\sigma^{B}_{3},
\eeq
where $\sigma_{3}=\frac{1}{2}(|e\rangle\langle e|-|g\rangle\langle g|)$ is the pseudospin operator. $H_{I}$ denotes the interaction Hamiltonian between the objects and gravitational fields,  which has the form
\beq
H_{I}=-\frac{1}{2}\big[Q_{A}^{ij}E_{ij}(\vec x_{A})+Q_{B}^{ij}E_{ij}(\vec x_{B})\big].
\eeq
Here $Q_{A(B)}^{ij}$ is the quadrupole moment of the object A (B),  which is induced by quantum gravitational fluctuations in vacuum, and $E_{ij}$ is the gravitoelectric tensor defined as $E_{ij}=R_{0i0j}$ which satisfies the linearized Einstein field equations arranged in a form in analogy to the Maxwell equations~\cite{WB}, where $R_{\mu\nu\tau\rho}$ is the Riemann tensor. The metric tensor for the fluctuating gravitational fields can be expanded  as $g_{\mu\nu}=\eta_{\mu\nu}+h_{\mu\nu}$, where $\eta_{\mu\nu}$ is the metric of the flat Minkowski spacetime, and $h_{\mu\nu}$ the linearized perturbations. So $E_{ij}$ can be expressed as
\beq\label{Eij}
E_{ij}=\frac{1}{2}\ddot h_{ij},
\eeq
where a dot denotes the derivative with respect to time $t$.  In the transverse tracefree (TT) gauge, the quantized $h_{ij}$ takes the standard form~\cite{TC}
\beq\label{hij}
h_{ij}=\sum_{\vec k,\lambda}\frac{1}{\sqrt{2\omega(2\pi)^3}}[a_{\lambda}(\vec k) e_{ij}(\vec k,\lambda) e^{i(\vec k\cdot \vec x-\omega t)}+H.c.]
\eeq
Here $H.c.$ denotes the Hermitian conjugate, $\omega=|\vec k|=(k_{x}^2+k_{y}^2+k_{z}^2)^{1/2}$, $a_{\lambda}(\vec k)$ is the gravitational field annihilation operator, i.e.  $a_{\lambda}(\vec k)|\{0\}\rangle=0$, $\lambda$ labels the polarization states, and $e_{ij}(\vec k,\lambda)$ are polarization tensors.

We assume that the two objects are prepared in the symmetric or antisymmetric state,
\beq\label{psi}
\psi_{\pm}=\frac{1}{\sqrt{2}}(|g_{A}\rangle|e_{B}\rangle \pm |e_{A}\rangle|g_{B}\rangle),
\eeq
both of which are maximally entangled. To investigate the induced quadrupole-quadrupole interaction energy between two entangled objects due to the fluctuating gravitational field in vacuum, we exploit the DDC formalism~\cite{JJ,JJC}, which allows us to identify the contributions of the \emph{vacuum fluctuations} and  \emph{radiation reaction}  to the interaction energy, respectively.  The effective Hamiltonian that governs the time evolution of object A's (or B's) observables~\cite{LM} is given by the summation of the following two terms
\beq
(H^{eff}_{A})_{vf}=-\frac{i}{8}\int^{t}_{t_{0}}d t' C^{F}_{ijkl}(x_{A}(t),x_{A}(t'))[Q_{AF}^{ij}(t),Q_{AF}^{kl}(t')],
\eeq
\bea
\nonumber (H^{eff}_{A})_{rr}=-\frac{i}{8}\int^{t}_{t_{0}}d t'&& \bigg[\chi^{F}_{ijkl}(x_{A}(t),x_{A}(t'))\{Q_{AF}^{ij}(t),Q_{AF}^{kl}(t')\} \\ &&+\chi^{F}_{ijkl}(x_{A}(t),x_{B}(t'))\{Q_{AF}^{ij}(t),Q_{BF}^{kl}(t')\} \bigg],
\eea
where $Q_{A(B)F}^{ij}$ denotes the free part of $Q_{A(B)}^{ij}$, which presents even in the absence of interaction. Similar expressions for object B can be obtained by exchanging the subscript A with B. Here we have introduced the statistical functions $C^{F}_{ijkl}(x_{A}(t),x_{A}(t'))$ and $\chi^{F}_{ijkl}(x_{A}(t),x_{A}(t'))$ for the gravitational field,
\beq
C^{F}_{ijkl}(x(t),x(t'))=\frac{1}{2}\langle \{0\}|\{E^{F}_{ij}(x(t)),E^{F}_{kl}(x(t')) \}|\{0\}\rangle,
\eeq
\beq\label{Fijkl}
\chi^{F}_{ijkl}(x(t),x(t'))=\frac{1}{2}\langle \{0\}|[E^{F}_{ij}(x(t)),E^{F}_{kl}(x(t'))]|\{0\}\rangle,
\eeq
where $E^{F}_{ij}$ denotes the free part of $E_{ij}$.

To obtain the energy shift for the system considered, we now evaluate the average values of the effective Hamiltonians $(H^{eff}_{A(B)})_{vf}$ and $(H^{eff}_{A(B)})_{rr}$ on the entangled state $(\ref{psi})$, and get
\beq\label{Evf}
(\delta E_{A})_{vf}=-\frac{i}{4}\int^{t}_{t_{0}}d t' C^{F}_{ijkl}(x_{A}(t),x_{A}(t'))\chi_{A}^{ijkl}(t,t'),
\eeq
\beq\label{Esr}
(\delta E_{A})_{rr}=-\frac{i}{4}\int^{t}_{t_{0}}d t' \left[\chi^{F}_{ijkl}(x_{A}(t),x_{A}(t'))C_{A}^{ijkl}(t,t') +\chi^{F}_{ijkl}(x_{A}(t),x_{B}(t'))C_{AB}^{ijkl}(t,t')\right],
\eeq
where $\chi_{A(B)}^{ijkl}(t,t')$ and $C_{A(B)}^{ijkl}(t,t')$ are respectively the antisymmetric and symmetric statistical functions of object A (B), while $\chi_{AB}^{ijkl}(t,t')$ and $C_{AB}^{ijkl}(t,t')$ are the collective statistical functions of the two-object system,
\beq
\chi_{AB}^{ijkl}(t,t')=\frac{1}{2}\langle\psi_{\pm}|[Q_{AF}^{ij}(t),Q_{BF}^{kl}(t')]|\psi_{\pm}\rangle,
\eeq
\beq
C_{AB}^{ijkl}(t,t')=\frac{1}{2}\langle\psi_{\pm}|\{Q_{AF}^{ij}(t),Q_{BF}^{kl}(t')\}|\psi_{\pm}\rangle.
\eeq
It is obvious that the contribution of vacuum fluctuations to the energy shift given in Eq. (\ref{Evf}) does not depend on the distance between the two objects, so it does not contribute to the interaction energy  directly.
Actually, only the second term on the right-hand side of Eq. (\ref{Esr}) depends on the distance between the two objects. Hence, the interaction between objects A and B comes solely from the contribution of the \emph{radiation reaction} (i.e. the radiation field produced by the quadrupole moment, which is induced vacuum fluctuations of  the quantized gravitational field). Thus, the interaction energy can be obtained as
\beq
\delta E_{AB}=-\frac{i}{4}\int^{t}_{t_{0}}d t'\left[\chi^{F}_{ijkl}(x_{A}(t),x_{B}(t'))C_{AB}^{ijkl}(t,t')+(A\rightleftharpoons B)\right].
\eeq
Apparently, the interaction originates from the exchange of a graviton between the two objects that are in a correlated state, and thus  the interaction energy between them is not  directly caused by vacuum fluctuations, but due to the field radiated by the induced quadrupole moment (radiation reaction).

\section{Resonance interaction between two entangled  gravitational polarizable objects}

To obtain the explicit expression of $\delta E_{AB}$, we now evaluate the statistical function of the gravitational field $\chi^{F}_{ijkl}(x_{A}(t),x_{B}(t'))$. According to Eqs. (\ref{Eij}), (\ref{hij}), and (\ref{Fijkl}), it is easy to get
\bea
\nonumber \chi^{F}_{ijkl}(x(t),x(t'))&=&\frac{1}{8}\langle \{0\}|[\ddot h_{ij}(x(t)),\ddot h_{kl}(x(t'))]|\{0\}\rangle\\
&=&\frac{1}{8(2\pi)^3}\int d^{3}\vec k \sum_{\lambda}e_{ij}(\vec k,\lambda)e_{kl}(\vec k,\lambda) \frac{\omega^{3}}{2}\left(e^{-i\omega\Delta t}-e^{i\omega\Delta t} \right)e^{i\vec k\cdot(\vec x-\vec x')},
\eea
where $\Delta t=t-t'$. Here the summation of polarization tensors in the $TT$ gauge gives~\cite{HL}
\bea
\nonumber\sum_{\lambda}e_{ij}(\vec k,\lambda)e_{kl}(\vec k,\lambda)=&&\delta_{ik}\delta_{j l}+\delta_{il}\delta_{j k}-\delta_{ij}\delta_{k l}+\hat k_{i}\hat k_{j}\hat k_{k}\hat k_{l}+\hat k_{i}\hat k_{j}\delta_{k l}\\
&&+\hat k_{k}\hat k_{l}\delta_{ij}-\hat k_{i}\hat k_{k}\delta_{j l}-\hat k_{i}\hat k_{l}\delta_{j k}-\hat k_{j}\hat k_{k}\delta_{i l}-\hat k_{j}\hat k_{l}\delta_{i k},
\eea
where
\beq
\hat k_{i}=\frac{\vec k_{i}}{|\vec k|}=\frac{\vec k_{i}}{\omega}.
\eeq
Transforming to the spherical coordinate, $\hat k_1=\sin{\theta}\cos{\phi}$, $\hat k_2=\sin{\theta}\sin{\phi}$, $\hat k_3=\cos{\theta}$, and letting
\beq
\sum_{\lambda}e_{ij}(\vec k,\lambda)e_{kl}(\vec k,\lambda)=g_{ijkl}(\theta,\phi),
\eeq
we obtain
\beq
\chi^{F}_{ijkl}(r,\Delta t)=\frac{1}{16(2\pi)^3}\int_{0}^{\infty}\omega^{5}\left(e^{-i\omega\Delta t}-e^{i\omega\Delta t} \right) d\omega \int_{0}^{\pi}\sin{\theta}d\theta \int_{0}^{2\pi} g_{ijkl}(\theta,\phi)e^{i\omega r\cos{\theta}}d \phi,
\eeq
where $r=|\Delta\vec x|=|\vec x-\vec x'|$ is the distance between the two objects. After some algebraic calculations, we obtain
\bea
\nonumber\chi^{F}_{1111}(r,\Delta t)=\frac{-i}{32\pi r^{5}}\bigg[&&\Big(r^{4}\partial^{4}_{t'}-2r^3\partial^{3}_{t'}+5r^2\partial^{2}_{t'}-9 r\partial_{t'}+9\Big)\delta(r-\Delta t)\\
&&-\Big(r^{4}\partial^{4}_{t'}+2r^3\partial^{3}_{t'}+5r^2\partial^{2}_{t'}+9 r\partial_{t'}+9\Big)\delta(r+\Delta t)\bigg],\\
\nonumber\chi^{F}_{1212}(r,\Delta t)=\frac{-i}{32\pi r^{5}}\bigg[&&\Big(r^{4}\partial^{4}_{t'}-2r^3\partial^{3}_{t'}+3r^2\partial^{2}_{t'}-3 r\partial_{t'}+3\Big)\delta(r-\Delta t)\\
&&-\Big(r^{4}\partial^{4}_{t'}+2r^3\partial^{3}_{t'}+3r^2\partial^{2}_{t'}+3 r\partial_{t'}+3\Big)\delta(r+\Delta t)\bigg],\\
\nonumber\chi^{F}_{1122}(r,\Delta t)=\frac{i}{32\pi r^{5}}\bigg[&&\Big(r^{4}\partial^{4}_{t'}-2r^3\partial^{3}_{t'}+r^2\partial^{2}_{t'}+3 r\partial_{t'}-3\Big)\delta(r-\Delta t)\\
&&-\Big(r^{4}\partial^{4}_{t'}+2r^3\partial^{3}_{t'}+r^2\partial^{2}_{t'}-3 r\partial_{t'}-3\Big)\delta(r+\Delta t)\bigg],\\
\nonumber\chi^{F}_{1313}(r,\Delta t)=\frac{i}{16\pi r^{5}}\bigg[&&\Big(-r^3\partial^{3}_{t'}+3r^2\partial^{2}_{t'}-6 r\partial_{t'}+6\Big)\delta(r-\Delta t)\\
&&-\Big(r^3\partial^{3}_{t'}+3r^2\partial^{2}_{t'}+6 r\partial_{t'}+6\Big)\delta(r+\Delta t)\bigg],
\eea
\beq
\nonumber\chi^{F}_{1133}(r,\Delta t)=\frac{i}{8\pi r^{5}}\bigg[\Big(r^2\partial^{2}_{t'}-3 r\partial_{t'}+3\Big)\delta(r-\Delta t)-\Big(r^2\partial^{2}_{t'}+3 r\partial_{t'}+3\Big)\delta(r+\Delta t)\bigg],
\eeq
\beq
\nonumber\chi^{F}_{3333}(r,\Delta t)=\frac{-i}{4\pi r^{5}}\bigg[\Big(r^2\partial^{2}_{t'}-3 r\partial_{t'}+3\Big)\delta(r-\Delta t)-\Big(r^2\partial^{2}_{t'}+3 r\partial_{t'}+3\Big)\delta(r+\Delta t)\bigg],
\eeq
\beq
\chi^{F}_{1111}=\chi^{F}_{2222},\quad\chi^{F}_{1212}=\chi^{F}_{1221}=\chi^{F}_{2121}=\chi^{F}_{2112},\quad \chi^{F}_{1122}=\chi^{F}_{2211},
\eeq
\beq
\chi^{F}_{1313}=\chi^{F}_{3113}=\chi^{F}_{1331}=\chi^{F}_{3131}=\chi^{F}_{2323}=\chi^{F}_{3223}=\chi^{F}_{2332}=\chi^{F}_{3232},
\eeq
\beq
\chi^{F}_{1133}=\chi^{F}_{3311}=\chi^{F}_{2233}=\chi^{F}_{3322},
\eeq
with other components being zero.

Similar to the electromagnetic field case~\cite{LM}, the statistical function $C_{AB}^{ijkl}(t,t')$ can be obtained as
\beq
C_{AB}^{ijkl}(t,t')=\pm\frac{1}{2}q_{A}^{ij}q_{B}^{kl}(e^{-i\omega_{0}\Delta t}+e^{i\omega_{0}\Delta t}),
\eeq
where the $\pm$ sign  refer to the symmetric state and the antisymmetric  state, respectively, $q_{A(B)}^{ij}=e^{i\omega_{0}t}\langle g_{A(B)}|Q_{A(B)F}^{ij}|e_{A(B)}\rangle$, $q_{A(B)}^{*ij}=e^{-i\omega_{0}t}\langle e_{A(B)}|Q_{A(B)F}^{ij}|g_{A(B)} \rangle$, and here we have assumed that $q_{A(B)}^{ij}=q_{A(B)}^{*ij}$.

Then, the interaction energy $\delta E_{AB}$  between the two objects can be expressed as
\beq
\delta E_{AB}=\pm\frac{1}{64\pi r^{5}}\sum_{ijkl}D_{ijkl}q_{A}^{ij}q_{B}^{kl},
\eeq
where we have introduced
\beq
D_{1111}=-\omega_{0}^{4}r^{4}\cos{\omega_{0}r}+2\omega_{0}^{3}r^{3}\sin{\omega_{0}r} +5\omega_{0}^{2}r^{2}\cos{\omega_{0}r}-9\omega_{0}r\sin{\omega_{0}r}-9\cos{\omega_{0}r},
\eeq
\beq
D_{1212}=-\omega_{0}^{4}r^{4}\cos{\omega_{0}r}+2\omega_{0}^{3}r^{3}\sin{\omega_{0}r} +3\omega_{0}^{2}r^{2}\cos{\omega_{0}r}-3\omega_{0}r\sin{\omega_{0}r}-3\cos{\omega_{0}r},
\eeq
\beq
D_{1122}=\omega_{0}^{4}r^{4}\cos{\omega_{0}r}-2\omega_{0}^{3}r^{3}\sin{\omega_{0}r}-\omega_{0}^{2}r^{2}\cos{\omega_{0}r} -3\omega_{0}r\sin{\omega_{0}r}-3\cos{\omega_{0}r},
\eeq
\beq
D_{1313}=-2\Big(\omega_{0}^{3}r^{3}\sin{\omega_{0}r}+3\omega_{0}^{2}r^{2}\cos{\omega_{0}r}-6\omega_{0}r\sin{\omega_{0}r}-6\cos{\omega_{0}r} \Big),
\eeq
\beq
D_{1133}=-4\Big(\omega_{0}^{2}r^{2}\cos{\omega_{0}r}-3\omega_{0}r \sin{\omega_{0}r}-3\cos{\omega_{0}r} \Big),
\eeq
\beq
D_{3333}=8\Big(\omega_{0}^{2}r^{2}\cos{\omega_{0}r}-3\omega_{0}r \sin{\omega_{0}r}-3\cos{\omega_{0}r} \Big),
\eeq
\beq
D_{1111}=D_{2222},\quad D_{1212}=D_{1221}=D_{2121}=D_{2112},\quad D_{1122}=D_{2211},
\eeq
\beq
D_{1313}=D_{3113}=D_{1331}=D_{3131}=D_{2323}=D_{3223}=D_{2332}=D_{3232},
\eeq
\beq
D_{1133}=D_{3311}=D_{2233}=D_{3322},
\eeq
with other components of $D_{ijkl}$ being zero. Recall that the quadrupole tensor is symmetric, i.e. $q_{A(B)}^{ij}=q_{A(B)}^{ji}$. In the near regime $r\ll \omega^{-1}_{0}$, the leading term of the interaction energy takes the form
\bea
\nonumber\delta E_{AB}\simeq\mp\frac{\cos{\omega_{0}r}}{64\pi r^5}\bigg[&&24q_{A}^{33}q_{B}^{33}-12\Big(q_{A}^{11}q_{B}^{33}+q_{A}^{33}q_{B}^{11}+q_{A}^{22}q_{B}^{33} +q_{A}^{33}q_{B}^{22}+4 q_{A}^{13}q_{B}^{13} +4 q_{A}^{23}q_{B}^{23} \Big)\\
&&+3\Big(q_{A}^{11}q_{B}^{22}+q_{A}^{22}q_{B}^{11}+4 q_{A}^{12}q_{B}^{12}\Big)+9\Big(q_{A}^{11}q_{B}^{11} +q_{A}^{22}q_{B}^{22}\Big)\bigg],
\eea
where the $\mp$ sign refer to the symmetric state and the antisymmetric state respectively. Since in the near regime $\cos{\omega_{0}r}\to 1$, the quantum gravitational interaction between two entangled objects varies with distance as $r^{-5}$. In the SI units, it can be rewritten as
\bea
\nonumber\delta E_{AB}\simeq\mp\frac{G\hbar\omega_{0}}{4 r^5}\bigg[&&24\alpha^{3333}-12\Big(\alpha^{1133}+\alpha^{3311}+\alpha^{2233} +\alpha^{3322}+4 \alpha^{1313} +4 \alpha^{2323} \Big)\\
&&+3\Big(\alpha^{1122}+\alpha^{2211}+4 \alpha^{1212}\Big)+9\Big(\alpha^{1111} +\alpha^{2222}\Big)\bigg],
\eea
if we define a gravitational polarizability $\alpha^{ijkl}\equiv q^{ij}q^{kl}/\hbar\omega_0$ in analogy to electrodynamics.
It is worth pointing out that  the gravitational polarizability defined here is  actually  $\hbar-$independent although it  carries $\hbar$ in the denominator.  The reason is that the element of the transition matrix of the quadrupole operator $q_{A(B)}^{ij}$ is $\hbar-$dependent. This is analogous to the electromagnetic field case where the electric polarizability of neutral atoms is similarly defined. We  define the gravitational polarizability in a such way so as to ensure its $\hbar-$independence since it is a classical physical quantity and it should be  $\hbar-$independent as is demonstrated in Ref.~\cite{LMJ} for a particular case.
Obviously, the quantum gravitational interaction in the short-distance can be attractive or repulsive depending on the polarization and whether the state is symmetric or antisymmetric. For example, if the objects are isotropically polarizable, i.e. $\alpha^{ijkl}=\alpha$, then, $\delta E_{AB}$ can be simplified to be
\beq
\delta E_{AB}\simeq\pm\frac{21G\hbar\omega_{0}\alpha}{ r^5}.
\eeq
That is, when the two objects are in the symmetric or antisymmetric state, the quantum gravitational interaction in the short-distance is repulsive or attractive respectively.
In the far regime, i.e., $r\gg \omega^{-1}_{0}$, we have
\bea \label{Eab}
\nonumber\delta E_{AB}\simeq\mp\frac{G\hbar\omega_{0}^{4}}{4 r^{2}c^{3}}&&\bigg[2\left(4\alpha^{1313}+4\alpha^{2323}-4\alpha^{1212} -\alpha^{1111}-\alpha^{2222}+\alpha^{1122}+\alpha^{2211} \right)\sin{\frac{\omega_{0}r}{c}}\\ &&+\frac{\omega_{0}r}{c}\Big(4\alpha^{1212}+\alpha^{1111} +\alpha^{2222}-\alpha^{1122}-\alpha^{2211}\Big)\cos{\frac{\omega_{0}r}{c}} \bigg].
\eea
For given objects, whether the quantum gravitational interaction is attractive or repulsive depends on the polarization, the state of the objects, and the distance between them. If we assume that the objects are isotropically polarizable, Eq. ($\ref{Eab}$) can be simplified as
\bea \label{Eab iso}
\delta E_{AB}\simeq\mp\frac{G\hbar\omega_{0}^{4}\alpha}{ r^{2}c^{3}}\bigg(2\sin{\frac{\omega_{0}r}{c}} +\frac{\omega_{0}r}{c}\cos{\frac{\omega_{0}r}{c}} \bigg)=\mp\frac{G\hbar\omega_{0}^{4}\alpha}{r^{2}c^{3}}\sqrt{4+\frac{\omega_0^2 r^2}{c^2}}\cos{\left(\frac{\omega_{0}r}{c}-\phi\right)},
\eea
where $\phi=\arcsin{\frac{2}{\sqrt{4+\omega_0^2 r^2/c^2}}}$. Since $r\gg\omega_0^{-1}c$ in the far regime, Eq. (\ref{Eab iso})  can be further simplified as
\beq\label{Eab sim}
\delta E_{AB}\simeq\mp\frac{G\hbar\omega_{0}^{5}\alpha}{r c^{4}}\cos{\left(\frac{\omega_{0}r}{c}-\phi\right)}.
\eeq
Therefore, the quantum gravitational interaction oscillates with a decreasing amplitude which is proportional to $r^{-1}$.

A few comments are now in order. First, compared to the gravitational quadrupole-quadrupole interaction  induced by quantum gravitational vacuum fluctuations between two gravitationally polarizable objects in their ground states, which is a fourth-order effect as shown in Ref.~\cite{PJ} based on the leading order perturbation theory, the quantum gravitational interaction for two entangled objects considered here is a second-order effect, which is much greater.

Second, we would like to compare the extremum  of the resonance quantum gravitational interaction $\delta E_{AB}$ with the monopole-monopole quantum gravitational interaction $V_{m}\sim \frac{\hbar G^{2}M^{2}}{r^{3}c^{3}}$~\cite{JF} in the far regime.
For an object (for example  one which can be treated gravitationally as an elastic sphere) with radius $R$, mass $M$ and frequency $\omega_{0}$, the gravitational quadrupole polarizability $\alpha\sim MR^{2}/\omega_{0}^{2}$~\cite{LMJ}.  Then, the ratio between the amplitude of $\delta E_{AB}$ (\ref{Eab sim}) and $V_{m}$ takes the form
\beq\label{Eab Vm}
\frac{\delta E_{AB}}{V_{m}}\sim\frac{R^{2}\omega_{0}^{3}r^{2}}{G M c}=\frac{R^{2}\omega_{0}c}{G M}\left(\frac{r}{\lambdabar}\right)^{2},
\eeq
where $\lambdabar=c/\omega_{0}$ is the reduced characteristic transition wavelength.
For a gravitationally bound system, the orbital frequency $\Omega =\sqrt{G M/R^{3}}$, which gives a lower bound on $\omega_0$ for any physical system \cite{LMJ}. Thus, Eq. (\ref{Eab Vm}) can be re-expressed as
\beq
\frac{\delta E_{AB}}{V_{m}}\gtrsim\frac{R^{2}\Omega c}{\sqrt{2} G M}\left(\frac{r}{\lambdabar}\right)^{2}=\left(\frac{R}{R_S} \right)^{\frac{1}{2}}\left(\frac{r}{\lambdabar}\right)^{2},
\eeq
where $R_S=2GM/c^2$ is the Schwarzschild radius. Obviously, $r\gg c/\omega_{0}=\lambdabar$ (in the far regime), and the radius of an object $R$ must be larger than its Schwarzschild radius $R_S$. Thus, in the far regime, the extreme value of the resonance quantum gravitational interaction (\ref{Eab iso}) is much greater than the monopole-monopole quantum gravitational interaction.

Third, as the resonance interaction energy may behave as $r^{-1}$ in the far regime, a question arises as to whether it can be comparable to the classical Newtonian potential. It is shown that the ratio between $\delta E_{AB}$ (consider the amplitude part of $\delta E_{AB}$ in Eq. (\ref{Eab sim})) and the classical Newtonian gravitational potential $V_{N}=GM^2/r$ can be approximated as
\beq\label{Eab Vn}
\frac{\delta E_{AB}}{V_{N}}\simeq\frac{\hbar R^{2} \omega_{0}^{3}}{M c^{4}}=\frac{\hbar \omega_{0}}{M c^{2}}\left( \frac{R}{\lambdabar}\right)^{2}.
\eeq
It is obvious that $\hbar\omega_{0} \ll M c^{2}$, i.e. the energy level spacing is much less than the energy corresponding to the mass. However, the radius $R$ can be both smaller or larger compared with the transition wavelength $\lambdabar$. For electrically bound objects, e.g. hydrogen atoms, the ratio (\ref{Eab Vn}) can be estimated as
\beq
\frac{\delta E_{AB}}{V_{N}}\sim\left(\frac{1~{\rm eV}}{1~{\rm GeV}}\right) \left(\frac{10^{-11}~{\rm m}}{{10^{-7}~{\rm m}}}\right)^{2} \sim10^{-17}.
\eeq
If the objects are bound gravitationally, the transition frequency is of the order of the orbital frequency $\Omega$. Then, we have
\beq
\frac{\delta E_{AB}}{V_{N}}\sim \left(\frac{l_P}{R}\right)^{2} \left(\frac{R_S}{R}\right)^{1/2},
\eeq
where $l_P=\sqrt{\hbar G/c^3}$ is the Planck length. Therefore, although the resonance quantum gravitational interaction can behave as $r^{-1}$ in the far regime, it is still very small compared with the classical Newtonian potential.

 \section{Discussion}
\label{sec_disc}
In this paper, we have investigated the resonance quantum gravitational quadrupole-quadrupole interaction between two entangled  quantum objects coupled with a bath of fluctuating  gravitational fields  in vacuum based on the DDC formalism and linearized quantum gravity.  Our result shows that the interaction energy between two entangled objects behaves as $r^{-5}$ in the near regime, and oscillates with a decreasing amplitude proportional to $ r^{-1}$ in the far regime. Compared to the gravitational quadrupole-quadrupole interaction  induced by quantum gravitational vacuum fluctuations between two gravitationally polarizable objects in their ground states, which is a fourth-order effect, the quantum gravitational interaction for the two entangled objects considered here is a second-order effect, which is much greater.   Moreover, the extremum of the resonance quantum gravitational quadrupole-quadrupole interaction  is also greater than the quantum gravitational monopole-monopole interaction in the far regime. In other words, the entanglement between the two objects significantly enhances the quantum gravitational interaction between them.  However, it is worth pointing out that the entangled state we considered here is not the only state that can enhance the quantum gravitational interaction, a direct product of the superposition of the eigenstates may also do.  Although the quantum gravitational interaction oscillates with an amplitude  behave as $r^{-1}$ in the far regime, an estimation shows that it cannot reach the Newtonian potential.

\begin{acknowledgments}

We appreciate very much the insightful comments and helpful suggestions by anonymous referees. This work was supported in part by the NSFC under Grants No. 11435006, No. 11690034, No. 11805063, and No. 11775077, and by the Science and Technology Innovation Plan of Hunan province under Grant No. 2017XK2019.

\end{acknowledgments}

\end{document}